# A Review on the Process of Automated Software Testing

Durga Shree N
Department of Software and Systems Engineering,
School of Information Technology and Engineering,
Vellore Institute of Technology, Vellore, India
durgashree.n2020@vitstudent.ac.in

Sree Dharinya S*
Department of Software and Systems Engineering,
School of Information Technology and Engineering,
Vellore Institute of Technology, Vellore, India
ssreedharinya@vit.ac.in
*Corresponding Author

Dasari Vijayasree
Department of Software and Systems Engineering,
School of Information Technology and Engineering,
Vellore Institute of Technology, Vellore, India
dasarivijayasree.2020@vitstudent.ac.in

Nadendla Sai Roopa
Department of Software and Systems Engineering,
School of Information Technology and Engineering,
Vellore Institute of Technology, Vellore, India
nadendlasai.roopa2020@vitstudent.ac.in

Anugu Arun
Department of Software and Systems Engineering,
School of Information Technology and Engineering,
Vellore Institute of Technology, Vellore, India
anugu.arun2020@vitstudent.ac.in

*Abstract*— **The requirements in automation, digitalization, and fast computations have loaded the IT sector with expectations of highly reliable, efficient, and cost-effective software. Given that the process of testing, verification, and validation of software products consumes 50-75% of the total revenue if the testing process is ineffective, "n" times the expenditure must be invested to mend the havoc caused. A delay in project completion is often attributed to the testing phase because of the numerous cycles of debugging process. The software testing process determines the face of the product released to the user. It sets the standard and reliability of a company's outputs. As the complexity increases, testing gets intense so as to examine all the outliers and various branches of the processing flow. The testing process is automated using software tools to avoid the tedious manual process of test input generation and validation criteria, which certifies the program only to a certain confidence level in the presence of outliers.**

## I. Introduction

Testing does not improve the quality of the program but helps in discerning quality holes. It reduces uncertainty and assures quality [19]. The process of Software Testing can be automated or implemented manually. These are capable of testing both open source and commercial software. Previous research has shown that professional programmers average six defects per 1000 lines of code [31]. Software tests allow us to compare the design and architecture with the functionality of software to learn the precision captured in the development process. Intensive tests are conducted to analyse the behaviour of various input spaces.

A study conducted by Microsoft indicated that it takes about 12 programming hours to manually locate and correct software defects. At this rate, it could take 11.4 years to debug a program of 350000 lines of code with costs in millions [31]. The results demonstrate the requirement for methods to test software effectively and efficiently. Testing is made less cumbersome, effective, and more efficient by implementing automated testing methods. Apart from speeding up the testing process, it also increases the amount of test coverage that can be done in a limited time resulting in higher quality software.

Automated software testing is one of the applications of software tools that verifies, validates, develops, delivers, and reviews the whole software product. The testing tools perform special automated testing for each test case of the software for which it suits.

*A. Testing*

The testing process involves planning, test case creation, execution of the test, validation, and debugging. A strong test strategy that focuses on testing new functionality in a dedicated testing environment is required for a successful test [19]. After cycles of revision are made on the software application, regression testing is done to ensure that the new functionalities and modifications have not adversely affected the previously existing code. The software is tested with both good and bad test cases to observe the reaction and stability of the software.

Generating test cases and expected outcomes for every possibility of execution is labour-intensive and time-consuming. Testing should continue for as long as the costs of finding and correcting defects are lower than the potential cost of a software failure after release to the end-user.

Testing is considered to be a phase that can be initiated after fundamental development initiation. But it is recommended to perform testing throughout the development process as the costs to correct defects increases rapidly throughout the development cycle. After every level of development, unit, system, integration, and acceptance, testing is initiated to execute debugging with minimal risk. Unit testing validates various modules individually. System testing is verifying each subsystem's interfaces and interdependencies separately, then merging them to test the system as a whole to check the fulfillment of functionalities specified in SRS. Integration testing incorporates all system components, including hardware, software, and human interaction, to guarantee that system requirements are met in actual or simulated scenarios. Acceptance testing is usually carried out by users and is performed using a BETA release of the software. The expense of repairing software problems after they have been released can be 100 times higher than the cost of preventing the same issue from developing in the first place [31].

*B. Automation*

Automated testing uses pre-programmed tools to maximize test coverage while minimizing the time and effort required to perform the test [19]. For projects that are behind time or over budget, automation cannot be employed as a solution. A whole development effort, including strategy and goal planning, establishing requirements, analysing alternatives, execution, and assessment, is required to automate a test. Automated testing tools are software products that can execute other software using test scripts. When a probable problem is discovered, a test script may be run again with the exact same inputs and sequence, something manual testing cannot ensure. Reuse of scripts saves time and helps ensure the stability of the software.

The standard of automation tools is determined by the process of implementation. Evaluation is expected to be done on the test system for its capability to run both software and automation tools simultaneously.

## II. Software Testing

A program's correctness can never be determined. Testing is done to detect faults that cause failure. Various types of testing are done to increase the confidence level of the performance of the software. Functional testing is a sort of software testing in which the system is put to test against its functional requirements. Usability testing is a sort of software testing that is done from the perspective of the end-user to see if the system is easy to use. Security testing is a sort of software testing that aims to find system vulnerabilities and ensure that the system's data and resources are safe from attackers. Performance testing is a sort of software testing that aims to establish how well a system performs in terms of responsiveness and stability when subjected to a specific load. Regression testing is a sort of software testing that aims to guarantee that changes to the program (enhancements or bug fixes) haven't had an undesirable effect on it. Compliance testing is a sort of testing that is used to assess whether a system complies with internal or external standards [20]. Depending on the type of software, various other customized tests are executed to ensure the working stability of the software.

Software Testing is broadly classified into:

*A. Static Testing*

Static test generation involves comprehending and analyzing the program code and using symbolic execution techniques to mimic abstract program executions in order to compute inputs to drive along defined execution paths or branches without actually executing the program [29].

*B. Dynamic Testing*

Each new input vector in a DART-directed search seeks to drive the program's execution along a different path. Such a directed search aims to push the program to sweep through all of its possible execution routes by repeating the procedure [32].

## III. Automated Testing

Software testing automation is the process of writing a program in any programming/scripting language that duplicates the manual test case steps with the help of an external automation helper tool. It involves creating toolkits to test the source code that has already been implemented. Its goal is to make the testing processes more automated. Both developing the program and writing test scripts are development tasks; the first is for the application itself, and the second is for the scripts that will be used to test the application.

Tests that can be automated:
1. Background processes, file logging, and database entry are examples of hard-to-reach areas of the system.
2. Frequently utilised features with a high risk of error: payment systems, registrations, and so on. Because crucial functioning tests take several minutes on average, automation assures quick errors.
3. Load tests, which examine a system's functioning in the face of a huge number of requests.
4. Template operations, such as data searches, the entry of multi-field forms, and the verification of their preservation.
5. Validation messages: erroneous data should be filled in the fields, and the validation should be checked.
6. End-to-end situations that take a long time.
7. Data verification including precise mathematical computations, such as accounting or analytical processing.
8. Verifying the accuracy of the data search. [20]

## IV. Manual and Automated Testing: A Comparison

Code visibility does not always affect test code coverage or fault detection rate in manual testing. Low code visibility, on the other hand, generally leads to low code coverage and a low fault detection rate in automated testing employing testing tools.

| Manual Testing | Automated Testing |
| --- | --- |
| The human tester executes use cases on a piece of software | Makes use of various tools to execute use cases |
| Time-consuming | Significantly faster |
| Allows exploratory and random testing | Does not allow for exploratory or random testing |
| Relatively smaller investment and ROI | Bigger investment and better ROI for the long term |
| Prone to errors because of human intervention | Highly robust and reliable |
| Easily adaptable to changes | Scripts require changes when UI changes are made |
| Requirement for human resources | Requirement for testing tools and automation engineers |
| Cost-effective for a smaller volume of testing | Cost-effective for large volume testing |
| Does not offer feasibility for performance testing | Allows load testing, stress testing and other performance tests |
| Suitable for AdHoc testing, exploratory testing and cases where there are frequent changes of previous tests. | Suitable for regression testing, load testing, highly repeatable functional test cases. |
| Test metrics:<br>• Test case productivity metrics<br>• Defect acceptance metrics<br>• Test estimation metrics<br>• Defect rejection metrics<br>• Bad fix defect metrics<br>• Test estimation productivity metrics<br>• Test execution efficiency metrics [5]. | Test metrics:<br>• Automation scripting productivity<br>• Automation test execution productivity<br>• Automation coverage metrics<br>• Cost comparison metrics |

Table 1. Manual vs Automated Testing

When compared to manual testing, the cost of automated testing is higher, especially at the start of the automation process. Software testing tools and hardware are both very expensive. After a period of time, the return on investment will be positive. When compared to manual testing, there is no cost when running automated tests, which implies a cost each time a test is run, depending on the size of the testing procedure.

The total cost of testing is defined as the sum of the costs of manual and automated testing:
CT = CM + CA +a, where 'a' denotes other costs [9].

Software testing cannot be completely automated. The automation of the testing process can reduce testing costs by reducing the time spent on creating and running test cases.

## V. AUTOMATED TESTING TOOLS AND FRAMEWORKS

### A. Automation Frameworks

The automation framework comprises a combination of tools and practices that are designed to help during software testing. It consists of physical structures which are used for test creation and implementation as well as logical interaction. The automation frameworks are Linear Automation, Modular Based, Library Architecture, Data-Driven, Keyword-Driven, and Hybrid Testing Frameworks [16].

The Software Automated Testing Framework (SAT) has the following goals:
1. Improving and expanding collaboration between scholars and practitioners, as automation frameworks aid in the transition from theory to practice.
2. Using both record/playback scripting techniques and programmable scripting techniques (easier initial creation) (ease of maintenance).
3. Simplifying the test script maintenance process, as the authors anticipate that semi-automated testing frameworks will continue to be valuable and will make it easier for testing engineers to automate web application testing.

### B. Automation Tools

There are no specific standards for software testing tools. It is essential that we make the correct choice of automation tools for testing. Software testing tools can be classified as functional, loading and management testing tools which are further utilized on open source and commercial software. Automated software testing tools can be compared based on attributes such as programming language support, browser support, license, operating system support, price, supported programming languages, etc. Testing of security, performance, correctness, and reliability can be done by using software testing tools efficiently. Selection of tool depends on the software and technology stack which is to be used, detailed testing requirements, skillsets available in the organization, and license cost of the tool [15].

Various categories of tools are used for specified tasks. These are customized to perform the function allocated for them. Unit testing tools, functional testing tools, code coverage tools, test management tools and performance testing tools are a few essential toolkits for any automated testing process. PHPUnit, NUnit, JMockit and JUnit are samples of Unit Testing Tools. Functional Testing Tools include Selenium, Test Studio, HP QuickTest Professional, Tricentis Tosca Testsuite, TestComplete, Waitr, etc. Clover, PITest, CodeCover, Atlassian, Cobertura are a few Code Coverage Tools. Some of the Test Management Tools are Test Link, Test Environment Toolkit, Test Manager and TETware. Performance Testing Tools such as HP LoadRunner, JMeter, Silk Performer, and Rational Performance Tester determine the functioning standard of the automation process [16].

Tools used in automatic software testing include capture/playback tools to record testing sessions for future reference, test cases in script files, coverage analysers to check the coverage of testing, test case generators that use object and data models with requirements, logical and complexity analysers, code instrumentation tools to gather data about the program while execution, defect tracking tools to track detected errors and their resolution status and tools for test management that manage and organize script files, test cases, test reports and test results [9].

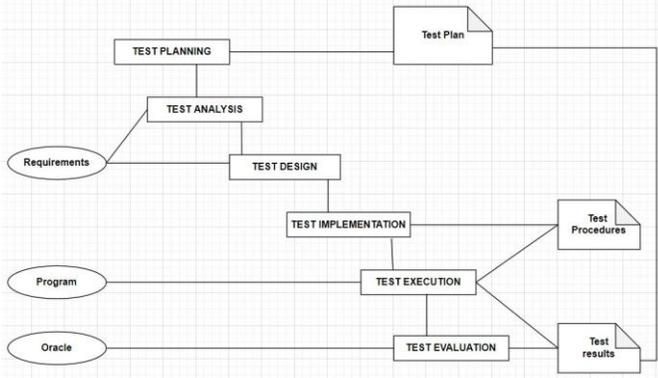

Fig. 1. Testing Process [9]

*C. Study of Tools and Algorithms*

A comparative study of automation testing tools, Testcomplete and quicktest pro is done in [23]. Recording efficiency, capability of generation of scripts, data-driven testing, test result reports, reusability, execution speed, ease of learning, cost and playback of the scripts are the parameters used to weigh the efficiency and effectiveness of testing tools.

Automatic test execution techniques are sensitive to changes during program implementation. A comprehensive testing language, TestTalk can be executed by different tools on different platforms to overcome the problem of maintaining port test descriptions. TestTalk test descriptions are tool-neutral and platform-independent. Various testing tools use different testing descriptions. TestTalk makes the software test descriptions represent a particular portion of a software project. The problems of being expensive to develop, sensitive to changes, and difficult to port can be dealt with using TestTalk. Changes in software implementation are sensitive to current automatic test execution methodologies [10].

An algorithm has been proposed in [17] to quantify dead code, exception, runtime errors and perform static analysis, unit testing and assertion-based testing. The algorithm is as follows:

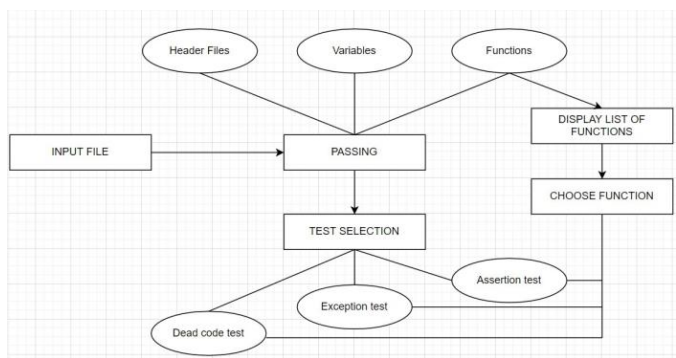

Fig. 2. The methodology proposed by [17]

*D. Benefits of Automated Testing Tools*

Identification of regressions, insulation from human errors, easy identification of bugs, shorter span of bug existence in code, testing code in live development, coordination between applications in play, incremental adoption to system evolution and availability of developer-oriented tools assure reliability on these tools. Automatic recovery from errors and enhanced fault tolerance ensures recoverability without heavy losses.

In order to ensure software quality, certain objectives are to be met: removal of maximum defects before product release, detection of errors at the earlier stage of SDLC and locating the source of defects. The crucial reason behind the use of automated software testing tools by the increased number of developers than ever before is the increasing complexity of software development.

## VI. VARIOUS APPROACHES TO AUTOMATED SOFTWARE TESTING

The quality of software tests depends upon the intensity of domain knowledge of the tester. Pre-planned tests are the bounds set on automation tests. Manual generation of test cases has a time constraint. So, each test case generated should be efficient enough to track down errors, non-redundant and of manageable complexity. Requirements determine the design of black-box testing; the clearer and more complete the requirements are, the better will be standard of test cases. By considering the requirement specifications as logical input-output data, cause-effect graphs and decision tables can be implemented to generate test cases automatically.

*A. The Data Mining Approach*

When testing new, potentially problematic releases of the system, the induced data mining models of tested software may be used to recover missing and incomplete requirements, construct a limited set of regression tests, and evaluate the quality of software outputs [1]. [33] has proposed Info-Fuzzy Network (IFN), a data mining algorithm to check the feasibility of the idea of data mining. This approach helps production of non-redundant test cases exploring almost all the existing functional relationships in a model. Since it does not rely on system code analysis like white-box techniques of automated test case selection, the method will be suited to testing complicated software systems. This procedure does not need a lot of human work. Missing, obsolete, or incomplete requirements are not an issue for the suggested approach, because it learns the functional links automatically from the execution data. Current test case creation approaches presume the availability of explicit requirements, which in many historical systems may be unavailable or inadequate.

*B. Automation Using Program Analysis*

Static code analysis supports the automation of code validation and inspection. Static code analysis tools are widely used because of automation possibility, scalability and numerous factor checkers. Techniques such as Fuzz testing which involves testing the program using the resulted data from randomly mutating well-defined and organized inputs, SAGE – a white-box fuzz testing that adopts a machine-code based approach for security testing, Pex – an automating unit testing for .NET, Yogi – combining testing and static analysis that simultaneously searches for both a test that acknowledges that the program satisfies the property and violates it are proposed in [6] that can generate inputs exhibiting each bug and predicates of input parameters responsible for it.

### C. Model-Checking Approach

Through the analysis of intermediate values of variables during program execution, white-box testing helps developers to discover whether or not a program is partially compatible with its defined behavior and design. The execution trace created by monitoring code placed into the program is frequently used to capture these intermediate values. The values in an execution trace are compared to the values expected by the stated behavior and design after the program has been completed. Inconsistences between expected and actual values can lead to the detection of specification and implementation mistakes. To resolve these conflicts and inconsistencies, [2] proposes an approach that (i) verifies execution traces using a model checker during white-box testing (ii) classifies and organizes execution traces based on requirement specifications into distinct equivalent partitions and (iii) applies popular model-checking counter-example generation mechanisms for unpopulated equivalence partitions.

### D. Hybrid Optimization Algorithm

A reliable product is to be delivered to the customer which requires assuring testing and validation procedures. Automation can help tune the product better. [8] provides an optimized automatic software testing model through differential evolution and ant colony optimization as a hybrid model to achieve improved accuracy and reliability in software testing. It is compared with other methods such as artificial neural networks and particle swarm optimization to prove better reliability and peak accuracy.

### E. DASE for Improving Automated Software Testing

Complexity in the process of software testing has been opening new doors to innovation in the testing and automation process. [12] The Document-Assisted Symbolic Execution (DASE) uses the concept of natural language processing for program documentation analysis to perform automatic input extraction and constrain generation for further symbolic execution of the testing process. To hunt for strategies and rank execution paths, this approach provides guidance based on semantic importance.

### F. Automated Software Test Data Generation

It is the process of identifying and generating test data justifying the required test criterion. Symbolic evolution is a widely used technique but involves a lot of algebraic complex manipulations. An alternative approach is proposed in [21] which is based on the actual execution of the program under test, function minimization methods and dynamic data flow analysis. Minimization of undesirable test case generation can be done by reducing the process of test data generation into a sequence of subgoals and these subgoals can be solved using minimization methods. Speeding up the search process to track the variables that potentially influence unexpected program behavior is done with the application of data flow analysis. The performance of the test data generation is least affected by the size of the input data.

## VII. Effectiveness and Efficiency of Automated Software Testing

Testing is automated to enhance the process of testing with better confidence levels on the correctness of the system developed. The idea of verification and validation performs the task of inspiring confidence in the integrity of the system. The two most important goals of software testing are: (i) Achieving in minimal time a given degree of confidence x in a program's correctness and (ii) Discovering a maximal number of errors within a given time bound, n [7]. The battle of choosing efficiency over effectiveness began as the requirement for assurance of software performance kept increasing leading to unbounded time and resources spent on validation. The most effective testing reveals a maximal number of errors and inspires a maximum degree of confidence in the correctness of a program. The most efficient testing technique (i) generates a sufficiently effective test suite in minimal time (ii) generates the most effective test suite in the given time budget.

Random testing generates partitions from input space where the partitions might not be mutually exclusive and exhaustive from the input set. Systematic testing generates input segments from analysis of source code and error-prone sections of the program. The efficiency of systematic testing decreases as the time spent on analysis increases while the efficiency of random testing remains unchanged [24]. Given any systematic testing technique S that discovers one partition for each input sampled, we introduce a hybrid technique H that starts with R and switches to S after a certain time to beat their drawbacks individually.

An efficient framework is essential to maintain and improve the effectiveness of software testing in the absence of program specifications that are valuable resources for elevating the effectiveness of software testing in generating test inputs and checking test executions for correctness. Extracting the redundancy concentration from existing tools to develop non-redundant test inputs that cover almost all the outliers and expose the behavioral differences during regression testing is implemented using the framework.

Manual test case generation and output specifications is a slow and extensive process. Test case generation tools can be utilized to find the differences between expected and actual outputs and analyze the behavioral differences after modifying a program. [24] suggests an efficient framework consisting of two major groups of components, one addressing the issues in generating test inputs such as the Redundant-test detector and non-redundant-test generator and the other addressing the issues in checking the correctness in execution and sending feedback information to the first group to guide test generation such as test selector, test abstractor and program-spectra comparator.

## VIII. APPLICATION, BENEFITS AND LIMITATIONS

### A. Automated Testing for Web Applications

Automation stands to include all testing tasks to overcome the drawbacks of traditional techniques using the Software Automated Testing (SAT) framework. It converts well-formatted test case steps into reusable programmable test scripts, which are then read by the Selenium automation tool and run in the web browser. Because the framework handles the source code generation step, there is no need for any programming experience to automate web applications. Adding or updating the freshly built testing steps is the only role of testers. When developing automation projects, eliminates the need to start from scratch. The primary goal of the proposed framework is to reduce the overall cost of the test automation process.

The suggested SAT architecture [13] can be thought of as a new layer that sits between testers and automation technologies. It's a desktop application that helps testers automate the process of writing test scripts. By automating this process, we can save a lot of time and money on testing, lowering the cost of the project. The scope of the proposed SAT framework is confined to web application regression software testing automation. There are five key processing processes in the basic framework engine.

The initialization stage involves the SAT framework crawling the SUT web page to identify all HTML controls that will be tested and ensuring that each HTMLcontrol has a predetermined list of values in the framework repository. The URL of the SUT should be provided to the framework by the tester. The framework then opens this URL and generates the TCS sheet and test script automatically. After both files have been generated, the tester can update them. Finally, the tester (or business user) can utilize the auto-generated script to execute the SUT.

### B. Automated testing for MATLAB

Unit testing can aid in improving the quality of science and engineering software, as well as how it is implemented. Unit tests should be automated if they are to be most successful, so that full test suites can be run fast and easily. The study in [14] shows the application of these concepts into practice using MATLAB, a testing framework. Using MATLAB xUnit, a test file that contains one or more test cases is created and the initial function in a test file always has the same shape. To make testing easy, the program's driver function runs tests to find all the test cases in test files, gathers them into a test suite, and displays the results. These test cases help identify off-by-one errors. MATLAB xUnit was built from the ground up to be used by procedural programmers, and its documentation is geared toward that purpose.

### C. Benefits and Limitations

In terms of various parameters affecting the automation process of software testing, we observe that the balance between these ensures benefits from the strategy. Improved product quality, test coverage, reduced testing time, reliability, increase in confidence, reusability of tests, less human effort, reduction in long-term cost and better ROI and increased fault detection are a few assets of automation. Manual testing is irreplaceable, especially those tests that require extensive knowledge in a domain. Difficulty in maintenance, sufficient time to gain maturity and unavailability of skillful people can become liabilities to automation software testing.

| Advantages | Disadvantages |
|---|---|
| Improves accuracy and quick finding of bugs compared to manual testing | Choosing the right tool requires considerable effort, time, and an evolution plan |
| Saves time and effort by making testing more efficient | Requires knowledge of the testing tool. |
| Increases test coverage because multiple testing tools can be used at once allowing for parallel testing of different test scenarios | The cost of buying the testing tool and, in the case of playback methods, test maintenance is a bit expensive |
| The automation test script is repeatable | Proficiency is required to write the automation test scripts |

Table 2. Advantages and Disadvantages of Automated Testing

## IX. CONCLUSION

Software test automation improves software quality and reduces software development costs in a long term by increasing the Return on Investment and further maturity of automatic test generation and debugging based on experience. A potential testing tool should be capable of exercising as many program statements as possible on execution when given a program statement with a set of input parameters. It is much needed to address the understandability, portability, and maintainability of executable test descriptions. The desire for repeatability and accuracy is one reason organizations are moving to automate testing, both to uncover bugs and ensure that they meet performance standards.